\documentclass[12pt]{article}
\usepackage{amssymb,amsmath,epsfig}
\allowdisplaybreaks
\begin{document}

\title{\bf Dynamics of Axial Symmetric System in Self-Interacting Brans-Dicke Gravity}

\author{M. Sharif $^1$ \thanks{msharif.math@pu.edu.pk} and Rubab
Manzoor $^2$\thanks{rubab.manzoor@umt.edu.pk}\\
$^1$ Department of Mathematics, University of the Punjab,\\
Quaid-e-Azam Campus, Lahore-54590, Pakistan.
\\$^2$ Department of Mathematics,\\ University of Management and
Technology,\\
Johar Town Campus, Lahore-54782, Pakistan.}
\date{}
\maketitle

\begin{abstract}
This paper investigates dynamics of axial reflection symmetric model
in self-interacting Brans-Dicke gravity for anisotropic fluid. We
formulate hydrodynamical equations and discuss oscillations using
time-dependent perturbation for both spin as well as
spin-independent cases. The expressions of frequency, total energy
density and equation of motion of oscillating model are obtained. We
study instability of oscillating models in weak approximations. It
is found that the oscillations and stability of the model depend
upon the dark energy source along with anisotropy and reflection
effects. We conclude that the axial reflection system remains stable
for stiffness parameter $\Gamma=1$, collapses for $\Gamma>1$ and
becomes unstable for $0<\Gamma<1$.
\end{abstract}

\section{Introduction}

Dark energy and stellar evolution are interesting issues of modern
cosmology as well as gravitational physics. Different astronomical
surveys (such as Sloan Digital Sky Survey, Wilkinson Microwave
Anisotropy Probe, Supernova type Ia, large scale-structure, weak
lensing and galactic cluster emission of X-rays etc.) reveal
accelerated expansion of the universe \cite{1}. It is assumed that a
mysterious form of energy termed as dark energy is responsible for
this accelerated expansion of the universe. The resolution of this
mystery leads to various modified theories of gravity by modifying
the Einstein-Hilbert action. In this context, the scalar-tensor
theory is one of the most fascinated idea which has provided
solutions of various cosmic problems such as early and late behavior
of the universe, inflation, coincidence problem and cosmic
acceleration \cite{2}.

The most explored and useful example of scalar-tensor gravitational
framework is the Brans-Dicke theory of gravity. This is a natural
generalization of general relativity constructed by the coupling of
tensor field $R$ and a massless scalar field $\phi$. It also
contains a constant tuneable parameter $\omega_{BD}$ which can be
tuned according to suitable observations. The concept of this theory
is based upon the weak equivalence principle, Mach's principle and
Dirac's large number hypothesis \cite{3}. The basic idea of this
theory is that the inertial mass of an object is not an intrinsic
property of the object itself but is generated by the gravitational
effect of all the other matter in the universe. For cosmic
inflation, this theory is generalized with self-interacting scalar
field by the inclusion of scalar potential function $V(\phi)$
\cite{4} known as self-interacting Brans-Dicke (SBD) gravity. This
has attracted a community of researchers for the viable discussion
of cosmic problems in scalar-tensor framework \cite{5, 5a}.

The study of formation and evolution of stars, galaxies and cluster
of galaxies has important implications in cosmology and
gravitational physics. Many observational and experimental surveys
such as Sloan Digital Sky Survey, University of Washington N-Body
shop, the Virgo consortium, Leiden observatory and the Hubble
telescope indicate stellar structures to resolve cosmic issues like
dark matter, dark energy and completeness of big-bang theory. It is
conventional that stellar models are mostly rotating and anisotropic
in nature. Anisotropy plays a significant role in different
dynamical phases of stellar evolutions \cite{6}.

The pattern of uniform as well as differential rotations of various
evolving celestial bodies are investigated through analysis of
stability and oscillations of axial configurations in weak
approximations. Arutyunyan et al. \cite{9} used Newtonian (N) and
post-Newtonian (pN) regimes to explore the structure of rotating
celestial object. Chandrasekhar and Friedman \cite{10} described
perturbation theory of axial symmetric models to discuss instability
ranges of uniformly rotating stars. Clifford \cite{10a} explored
oscillations and stability of differentially rotating axial
symmetric system. Sharif and Bhatti \cite{11} discussed reflection
symmetric axial non-static models and found that instability ranges
depend upon the stiffness parameter and also the spinning models are
more stable.

Many people \cite{5a,12} investigated stellar evolutions in the
modified theories. Since the evolution of such models passes through
different dynamical stages, this study can lead to correct theory of
gravity or it may reveal some modifications hidden in the structure
formation of the universe. In this paper, we explore dynamics of
non-static axial reflection model in the framework of SBD gravity
and study stellar evolution under Mach's principle. The paper is
organized in the following format. In the next section, we review
SBD gravity and axial system with reflection symmetry as well as
anisotropic fluid. Section \textbf{3} describes dynamical picture of
evolving axial system such as hydrodynamics, oscillations and
instability regimes. Final section summarizes the results.

\section{Self-Interacting Brans-Dicke Theory}

The SBD theory is represented by the following action \cite{4}
\begin{equation}\label{1}
S=\int d^{4}x\sqrt{-g} [\phi
R-\frac{\omega_{BD}}{\phi}\nabla^{\mu}{\phi}\nabla_{\mu}{\phi}-V(\phi)+\emph{L}_{m}],
\end{equation}
where $\emph{L}_{m}$ shows matter distribution and $8\pi G_0=c=1$.
Varying the above action with respect to $g_{\mu\nu}$ and $\phi$, we
obtain
\begin{eqnarray}\label{2}
G_{\mu\nu}&=&\frac{1}{\phi}(T_{\mu\nu}^{m}+T_{\mu\nu}^{\phi}),\\\label{3}
\Box\phi&=&\frac{T^{m}}{3+2\omega_{BD}}
+\frac{1}{3+2\omega_{BD}}[\phi\frac{dV(\phi)}{d\phi}-2V(\phi)].
\end{eqnarray}
Here $G_{\mu\nu}$ represents the Einstein tensor,
$\frac{T_{\mu\nu}^{m}}{\phi}$ indicates the contribution of matter
in the presence of scalar field,
$T^{m}=g^{\mu\nu}T_{\mu\nu}$ and $\square$ is the d'Alembertian
operator. The energy contribution due to scalar field is described
by
\begin{equation}\label{4}
T^{\phi}_{\mu\nu}=\phi_{,\mu;\nu}
-g_{\mu\nu}\Box\phi+\frac{\omega_{BD}}{\phi}[\phi_{,\mu}\phi_{,\nu}
-\frac{1}{2}g_{\mu\nu}\phi_{,\alpha}\phi^{,\alpha}]
-\frac{V(\phi)}{2}g_{\mu\nu},
\end{equation}
which is energy momentum tensor associated with Machian terms that
describes the interaction of scalar field with the geometry of the
distant matter distributions in the universe and the effects of its
potentials upon them.

Equations (\ref{2}) and (\ref{3}) represent SBD field equations and
SBD wave equation, respectively. The right hand side of Eq.(\ref{2})
indicates that both terms are sources of geometry (gravitation).
There also exists static field in axial symmetric SBD model which
has $\phi=\phi(t_{0})=constant$ with respect to cosmic time $t_{0}$
and generalizes Einstein equations with an effective cosmological
constant $V(\phi_{0})$ \cite{z}. These static field configurations
lead to the dynamics of non-static axial system.

In order to discuss dynamics of non-static axial symmetric
configurations, we consider non-static axially symmetric spacetime
characterized with reflection \cite{11, 13}
\begin{equation}\label{5}
ds^2=-A^2(t,r,\theta)dt^{2}+B^2(t,r,\theta)(dr^{2}+r^{2}d\theta^{2})+
2L(t,r,\theta)dtd\theta+C^2(t,r,\theta)d\phi^{2},
\end{equation}
having matter contribution in the form of locally anisotropic fluid
given by
\begin{equation}\label{7}
T_{\mu\nu}^{m}=(\rho+p)u_{\mu}u_{\nu}+pg_{\mu\nu}+\Pi_{\mu\nu}.
\end{equation}
Here
\begin{eqnarray}\nonumber
\Pi_{\mu\nu}&=&\frac{1}{3}(\Pi_{II}+2\Pi_{I})
(k_{\mu}k_{\nu}-\frac{1}{3}h_{\mu\nu})
+\frac{1}{3}(\Pi_{I}+2\Pi_{II})(\chi_{\mu}
\chi_{\nu}-\frac{1}{3}h_{\mu\nu})\\\nonumber
&+&\Pi_{k\chi}(k_{\mu}\chi_{\nu}+k_{\nu}\chi_{\mu}),
\end{eqnarray}
with
\begin{eqnarray}\nonumber
&&h_{\mu\nu}=g_{\mu\nu}+u_{\mu}u_{\nu},\quad
\Pi_{k\chi}=k^{\mu}\chi^{\nu}T_{\mu\nu},\quad
p=\frac{1}{3}h^{\mu\nu}T_{\mu\nu},\\\nonumber
&&\Pi_{I}=(2k^{\mu}k^{\nu}
-s^{\mu}s^{\nu}-\chi^{\mu}\chi^{\nu})T_{\mu\nu},\quad
\Pi_{II}=(2\chi^{\mu}\chi^{\nu}-k^{\mu}k^{\nu}-s^{\mu}s^{\nu})T_{\mu\nu},
\end{eqnarray}
where $\rho$ is the energy density, $p$ shows the isotropic
pressure, $\Pi_{\mu\nu}$ represents anisotropic stress tensor,
$\Pi_{I}\neq\Pi_{II}\neq\Pi_{k\chi}$ are the anisotropic scalars and
$h_{\mu\nu}$ expresses  projection tensor. The four velocity
$u_{\mu}$, unit four-vectors $k_{\mu},~s_{\mu}$ and $\chi_{\mu}$ are
calculated as
\begin{equation}\label{8}
u_{\mu}=-A\delta_{\mu}^{0}+\frac{L}{A}\delta^{2}_{\mu},\quad
k_{\mu}=B\delta_{\mu}^{2},\quad s_{\mu}=C\delta_{\mu}^{3},\quad
\chi_{\mu}=\frac{(\Delta)^{1/2}}{A}\delta^{2}_{\mu},
\end{equation}
with $\Delta=r^{2}A^{2}B^{2}+L^{2}$ and satisfy the following
relations
\begin{eqnarray}\nonumber
-u^{\mu}u_{\mu}=s^{\mu}s_{\mu}=k^{\mu}k_{\mu}=\chi^{\mu}\chi_{\mu}=1,\\
s^{\mu}u_{\mu}=k^{\mu}u_{\mu}=\chi^{\mu}u_{\mu}=s^{\mu}k_{\mu}
=\chi^{\mu}k_{\mu}=s^{\mu}\chi_{\mu}=0.
\end{eqnarray}
The non-zero components of the energy-momentum tensor due to scalar
field can be represented as
\begin{eqnarray}\label{9}
\frac{T^{\mu\nu({\phi})}}{{\phi}}=\left(
\begin{array}{cccc}
v_{1}+w_{1} & x_{1}+y_{1} & x_{3}+y_{3} & 0 \\
x_{1}+y_{1} &  v_{2}+w_{2} &  x_{2}+y_{2} & 0 \\
x_{3}+y_{3} &  x_{2}+y_{2} &  v_{3}+w_{3} & 0 \\
0 & 0 & 0 &  v_{4}+w_{4} \\
\end{array}
\right).
\end{eqnarray}
Here $v_{i}+w_{i}$ represents diagonal and $x_{j}+y_{j}$ shows
non-diagonal components of stress tensor (\ref{4}) in which $w_{i}$
and $y_{j}$ indicate axial reflection effects due to scalar field.

\section{Dynamics}

In this section, we carry out dynamical analysis of axial reflection
symmetric system. For this purpose, we derive the hydrodynamical
equations and discuss oscillations as well as instability ranges of
the perturbed axial system.

\subsection{Hydrodynamics}

The dynamical equations representing hydrodynamics of axially
symmetric system can be obtained with the help of Bianchi identity
$G^{\mu\nu}_{~~;\nu}=0$. This identity along with Eqs.(\ref{2}),
(\ref{4}) and (\ref{7}) provide the following equations for
$\mu=0,1,2,$
\begin{eqnarray}\nonumber
&&\dot{\rho}_{(m\phi)}-\rho_{(m\phi)}\left[\frac{\dot{B}}{B}
+\frac{\dot{C}}{C}+\frac{r^{2}A\dot{A}B^{2}}{\Delta}
+\frac{L\dot{L}}{\Delta}+\frac{r^{2}A^{2}B\dot{B}}{\Delta}\right]
+(\rho_{(m\phi)}+p_{(m\phi)})\\\nonumber
&&\times\frac{AB^{2}}{\Delta} \left[\frac{2r^{2}\dot{B}}{B}
+\frac{r^{2}\dot{C}}{C}+\frac{L^{2}\dot{B}}{A^{2}B^{3}}
-\frac{L^{2}\dot{A}}{A^{3}B^{2}}
+\frac{L\dot{L}}{A^{2}B^{2}}+\frac{L^{2}\dot{C}}{A^{2}B^{2}C}\right]\\\nonumber
&&+\frac{\Pi_{I(m\phi)}}{3\Delta}\left(\frac{\dot{B}}{B}-\frac{\dot{C}}{C}\right)
+\frac{\Pi_{II(m\phi)}}{\Delta}\left[r^{2}A^{2}B^{2}\left(\frac{\dot{B}}{B}-\frac{\dot{C}}{C}\right)
+\frac{L^{2}\dot{L}}{AL}
-\frac{L^{2}\dot{A}}{A^{2}}\right.\\\label{10}&&\left.-\frac{L^{2}\dot{C}}{AC}\right]
+E_{0}(t,r,\theta)=0,\\\nonumber
&&p'_{(m\phi)}+\frac{2}{9}\left(2\Pi'_{I(m\phi)}+\Pi'_{II(m\phi)}\right)
+\left[p_{(m\phi)}+\frac{2}{9}(2\Pi_{I(m\phi)}+\Pi_{II(m\phi)})\right]\\\nonumber
&&\times\left[\frac{C'}{C}+\frac{3LL'}{2}+\frac{r^{2}AA'B^{2}}{\Delta}
+\frac{r^{2}A^{2}BB'}{\Delta}+\frac{2rA^{2}B^{2}}{\Delta}
-\frac{rA^{2}B(rB)'}{\Delta}\right]\\\nonumber
&&-\frac{r^{2}AB^{5}}{(\Delta)^{3/2}}\left[\Pi^{\theta}_{k\chi(m\phi)}
-\Pi_{k\chi(m\phi)}\left(\frac{A^{\theta}}{A}
-\frac{6B^{\theta}}{B}-\frac{C^{\theta}}{C}-\frac{4r^{2}A^{2}B^{2}}{\Delta}
\left(\frac{A^{\theta}}{A}+\frac{B^{\theta}}{B}\right)\right.\right.\\\nonumber
&&\left.\left.-\frac{4LL^{\theta}}{\Delta}\right)\right]
+\frac{\rho_{(m\phi)}r^{4}A^{4}B^{4}}{\Delta^{2}}\left(B\dot{B}+\frac{A'}{A}
-\frac{LA^{\theta}}{r^{2}AB^{2}}\right)
-\frac{\rho_{(m\phi)}r^{2}A^{2}L^{2}B^{2}}{\Delta^{2}}\\\label{11}
&&\times\left(\frac{(rB)'}{rB}+\frac{L}{2L'}\right)
+E_{1}(t,r,\theta)=0,\\\nonumber
&&\frac{\rho_{(m\phi)}A^{2}B^{2}L}{\Delta^{2}}
\left[\frac{r^{2}\dot{\rho}_{(m\phi)}}{\rho_{(m\phi)}}+\frac{r^{2}\dot{A}}{A}
+3\frac{r^{2}\dot{B}}{B}+\frac{r^{2}\dot{L}}{L}
+\frac{r^{2}\dot{C}}{C}+\frac{1}{B^{2}}
\left(\frac{\rho^{\theta}_{(m\phi)}}{\rho_{(m\phi)}}\right.\right.\\\nonumber
&&\left.\left. +\frac{2L^{\theta}}{L}
+\frac{2A^{\theta}}{A}\right)+\frac{1}{\Delta}4r^{5}A^{2}
\left(\frac{\dot{A}}{A}+\frac{\dot{B}}{B}\right)
-\frac{4\dot{L}}{L\Delta}-\frac{5LAA^{\theta}}{\Delta}
-\frac{LA^{2}B^{\theta}}{B\Delta}\right.\\\nonumber
&&\left.+\frac{r^{2}A^{2}B^{2}}{\Delta}
\left(\frac{\dot{L}}{L}+\frac{\dot{B}}{B}\right)
+\frac{r^{2}A^{3}B^{2}A^{\theta}}{L\Delta}
-\frac{4L^{2}L^{\theta}\Delta}{r^{2}B^{2}}\right]
+\frac{\rho_{(m\phi)}A^{2}L^{2}}{\Delta^{2}}
\left(\frac{B^{\theta}}{B}+\frac{C^{\theta}}{C}\right.\\\nonumber
&&\left.-\frac{r^{2}BL\dot{B}}{\Delta}\right)
-\frac{r^{3}AB^{3}\Pi_{k\chi(m\phi)}}{\Delta^{3/2}}
\left[\frac{\Pi'_{k\chi(m\phi)}}{\Pi_{k\chi(m\phi)}}
+\frac{3}{r}+\frac{4B'}{B}+\frac{A'}{A}
+\frac{C'}{C}+\frac{3LL'}{2\Delta}\right.\\\nonumber
&&\left.+\frac{3r^{2}A^{2}B^{2}}{\Delta}\left(\frac{3}{r}
+\frac{2A'}{A}+\frac{3B'}{B}\right)+\frac{7LL'}{2\Delta}\right]
+\frac{1}{\Delta}\left(p_{(m\phi)}+\frac{2}{9}
\left(\Pi_{I(m\phi)}\right.\right.\\\nonumber
&&\left.\left.+2\Pi_{II(m\phi)}\right)\right)
\frac{1}{\Delta}\left[\frac{r^2A^2B^2}{\Delta}\left\{(2A^2+A)\left(
\frac{A^\theta}{A}\right.\right.\right.\left.+\frac{B^\theta}{B}\right)\\\nonumber
&&-L\frac{\dot{B}}{B}+\frac{2AB^\theta}{B}+2AA^\theta
+\frac{A^2C^\theta}{C}-\frac{r^2BL\dot{B}}{\Delta}
\left.-\frac{2A^2LL^\theta}{\Delta}
-\frac{L\dot{B}}{B}\right]-\frac{p}{C\Delta}\\\nonumber
&&\times(L\dot{C}+A^2C^\theta)+\frac{A^2}{\Delta}
\left\{p^\theta+\frac{2}{9}(\Pi^{\theta}_{I,(m\phi)}
+2\Pi^{\theta}_{II,(m\phi)})\right\}+E_{2}(t,r,\theta)=0.\\\label{12}
\end{eqnarray}
Here dot, prime and superscript $\theta$ indicate derivatives with
respect to time, $r$ and $\theta$, respectively. The subscript
$(m\phi)$ implies energy-momentum terms of matter distribution per
scalar field $(\frac{T^{\mu\nu(m)}}{\phi})$ which corresponds to
contributions of matter dynamics in the presence of scalar field.
The terms $E_{0}(t,r,\theta),~E_{1}(t,r,\theta)$ and
$E_{2}(t,r,\theta)$ represent energy contributions due to scalar
field and their values are given in Eqs.(\ref{A1})-(\ref{A3}).
Equations (\ref{10})-(\ref{12}) describe hydrodynamical equations of
axial reflection symmetric fluid in SBD gravity.

\subsection{Oscillations}

Now we discuss oscillations of axial system through perturbation
approach. We consider that initially the system is in hydrostatic
equilibrium and after that all metric functions along with dynamical
variables are perturbed with time dependence perturbation
$T(t)=e^{iw t}$ and the system starts oscillating with frequency
$w$. The metric tensor as well as the scalar field and scalar
potential have the same time dependence, while the dynamical
variables bear the same time dependence as follows
\begin{eqnarray}\label{13}
A(t,r,\theta)&=&A_0(r,\theta)+\epsilon e^{iw
t}a(r,\theta),\\\label{14} B(t,r,\theta)&=&B_0(r,\theta)+\epsilon
e^{iw t}b(r,\theta),\\\label{15}
C(t,r,\theta)&=&C_0(r,\theta)+\epsilon e^{iw
t}c(r,\theta),\\\label{16} L(t,r,\theta)&=&L_0(r,\theta)+\epsilon
e^{iw t}l(r,\theta),\\\label{16a}
\phi(t,r,\theta)&=&\phi_{o}(r,\theta)+\epsilon e^{iw
t}\Phi(r,\theta),\\\label{17} V(\phi)&=&V_0(r,\theta)+\epsilon e^{iw
t}{\bar{V}(r,\theta)},\\\label{17a} p(t,r,\theta)&=&p_{0}
(r,\theta)+\epsilon {\bar{p}}(iw t,r,\theta),\\\label{18}
\rho(t,r,\theta)&=&\rho_0(r,\theta)+\epsilon{\bar{\rho}}(iw
t,r,\theta),\\\label{19} \Pi_{I}(t,r,\theta)&=&\Pi_{I0}(r,\theta)+
\epsilon {\bar{\Pi}_{I}(iwt,r,\theta)},\\\label{20}
\Pi_{II}(t,r,\theta)&=&\Pi_{II0}(r,\theta)+ \epsilon
{\bar{\Pi}_{II}(iwt,r,\theta)},\\\label{21}
\Pi_{k\chi}(t,r,\theta)&=&\Pi_{k\chi0}(r,\theta)+ \epsilon
{\bar{\Pi}_{k\chi}(iwt,r,\theta)}.
\end{eqnarray}
Here $0<\epsilon\ll1$ and the subscript zero indicates static
distribution while terms having bar represent perturbed terms
\cite{11,14}.

Using Eqs.(\ref{13})-(\ref{21}), the perturbed configuration of
$02$-components of the field equations (\ref{2}) can be represented
as
\begin{equation}\nonumber
(lw^{2}+mw+n)e^{iw t}=0,
\end{equation}
where values of $l,~m$ and $n$ are given in (\ref{A4})-(\ref{A6}).
Since $e^{iwt}\neq0$, this implies that
\begin{equation}\label{23}
w=\frac{-m+\sqrt{m^{2}-4ln}}{2l},
\end{equation}
yielding frequency of the oscillating axial reflection system. This
shows that the frequency of oscillations depend upon the DE source
(scalar field), anisotropic effects and reflection configuration.

The total density of oscillating system can be obtained from the
perturbed form of the first law of conservation (\ref{10}) as
follows
\begin{eqnarray}\label{24}
\bar{\rho}_{(m\phi)}&=&\left[(F_{(m\phi)}
+\bar{E}_{0(a)})iw+\bar{E}_{0(b)}\right]e^{iwt}.
\end{eqnarray}
Here $F_{(m\phi)}$ shows contribution of matter with scalar field,
$\bar{E}_{0(a)}$ and $\bar{E}_{0(b)}$ represent scalar field
distributions whose values are given in Eqs.(\ref{A7})-(\ref{A8}).
The terms with subscript $(a)$ represents scalar field coupled to
frequency while subscript $(b)$ shows scalar field without
frequency. The perturbed form of Eq.(\ref{11}) provides the equation
of motion of the oscillating system given by
\begin{align}\nonumber
&\frac{1}{B_0^2}\left\{\bar{p}_{(m\phi)}'+\frac{2}{9}(2\bar{\Pi}_{I(m\phi)}'
+\bar{\Pi}_{II(m\phi)}')\right\}+\frac{1}{B_0^2}
\left\{\bar{p}_{(m\phi)}+\frac{2}{9}(2\bar{\Pi}_{I(m\phi)}+\bar{\Pi}_{II(m\phi)})
\right\}\\\nonumber &\times\left\{\frac{C_0'}{C_0}
+\frac{3L_0L_0'}{2}
+\frac{r^2A_0^2B_0^2}{\Delta_0}\left(\frac{A_0'}{A_0}
+\frac{B_0'}{B_0}+\frac{2}{r}
-\frac{1}{r}-\frac{B_0'}{B_0}\right)\right\}-\frac{r^2A_0B_0^3}
{\Delta^{\frac{3}{2}}_0}\bar{\Pi}_{k\chi(m\phi)}^{\theta}\\\nonumber
&-\bar{\Pi}_{k\chi(m\phi)}\frac{r^2A_0B_0^3}{\Delta_0^{\frac{3}{2}}}
\left\{\frac{A_{0}^{\theta}}{A_0}
+\frac{6B_{0}^{\theta}}{B_0}+\frac{C_{0}^{\theta}}{C_0}+\frac{4L_0L_{0}^{\theta}}
{\Delta_0}+\frac{4r^2A^2_0B^2_0}{\Delta_0}\left(\frac{A_{0}^{\theta}}{A_0}
+\frac{B_{0}^{\theta}}{B_0}\right)\right\}\\\nonumber
&+\frac{\bar{\rho}_{(m\phi)}r^4A_0^4}{\Delta_0^2}
\left(\frac{A_0'}{A_0}-\frac{L_0A_{0}^{\theta}}{r^2A_0B_0^2}\right)
-\bar{\rho}_{(m\phi)}L_0^2r^2\frac{A_0^2}{\Delta_0^2}
\left\{\frac{L_0}{2}+\frac{1}{r}+\frac{B_0'}{B_0}\right\}-e^{iwt}\left[
\frac{2b}{B_0^3}\right.\\\nonumber
&\times\left\{p_{0(m\phi)}'+\frac{2}{9}
(2{\Pi}_{I0(m\phi)}'+{\Pi}_{II0(m\phi)}')\right\}-
\left[\left\{\left(\frac{c}{C_0}\right)'+\frac{3L_0L_0'}{2}
\left(\frac{l}{L_0}+\frac{l'}{L_0'}\right)\right\}\right.\\\nonumber
&+\frac{r^2A_0^2B_0^2}{\Delta_0^2}\left(\frac{2a}{A_0}+\frac{2b}{B_0}
-\frac{\Delta_{p}}{\Delta_0}\right)
\left(\frac{A_0'}{A_0}+\frac{B_0'}{B_0}+\frac{2}{r}
-\frac{1}{r}-\frac{B_0'}{B_0}\right)+\frac{r^2A_0^2B_0^2}{\Delta_0}\\\nonumber
&\times\left.\left\{\left(
\frac{a}{A_0}+\frac{2b}{B_0}\right)'-\left(\frac{b}{B_0}\right)'\right\}\right]
\left\{p_{0(m\phi)}+\frac{2}{9}(2{\Pi}_{I0(m\phi)}+{\Pi}_{II0(m\phi)})\right\}
\frac{1}{B_0^2}-\frac{2b}{B_0^2}\\\nonumber
&\times\left\{p_{0(m\phi)}+\frac{2}{9}(2{\Pi}_{I0(m\phi)}+{\Pi}_{II0(m\phi)})
\right\}\left\{\frac{C_0'}{C_0}
+\frac{3L_0L_0'}{2}+\frac{r^2A_0^2B_0^2}{\Delta_0}\left(\frac{A_0'}{A_0}
+\frac{B_0'}{B_0}\right.\right.\\\nonumber &\left.\left.+\frac{2}{r}
-\frac{1}{r}-\frac{B_0'}{B_0}\right)\right\}+\frac{r^2A_0B_0^3}
{\Delta^{\frac{3}{2}}_0}{\Pi}_{k\chi0(m\phi)}^{\theta}
\left(\frac{a}{A_0}+\frac{3b}{B_0}-\frac{3\Delta_{p}}{\Delta_0}\right)
+\frac{r^3A_0B_0^3\Pi_{k\chi0(m\phi)}}{\Delta_0^{\frac{3}{2}}}\\\nonumber
&\times\left(\frac{a}{A_0}+\frac{3b}{B_0}-\frac{3{\Delta_{p}}}{\Delta_0}
\right)\left\{\frac{A_{0}^{\theta}}{A_0}
+\frac{6B_{0}^{\theta}}{B_0}+\frac{C_{0}^{\theta}}{C_0}+\frac{4L_0L_{0}^{\theta}}
{\Delta_0}+\frac{4r^2A^2_0B^2_0}{\Delta_0}\left(\frac{A_{0}^{\theta}}{A_0}
+\frac{B_{0}^{\theta}}{B_0}\right)\right\}\\\nonumber
&+\frac{r^3A_0B_0^3}{\Delta^{\frac{3}{2}}_0}{\Pi}_{k\chi0(m\phi)}
\left[\frac{6B_{0}^{\theta}}{B_0}\left(\frac{b^{\theta}}{B_{0}^{\theta}}
+\frac{b}{B_0}\right)\right.\left(\frac{a}{A_0}+\frac{c}{C_0}\right)^{\theta}
+\frac{4L_0L_{0}^{\theta}}{\Delta_0}\left(\frac{l}{L_0}
+\frac{l^{\theta}}{L_{0}^{\theta}} \right.\\\nonumber
&\left.\left.-\frac{\Delta_{p}}{\Delta_0}\right)
+\frac{4r^2A_0^2B_0^2}{\Delta_0}\left(\frac{2a}{A_0}+\frac{2b}{B_0}
-\frac{\Delta_{p}}{\Delta_0}\right)\left(\frac{a}{A_0}
+\frac{b}{B_0}\right)^{\theta}\right]-\frac{\rho_{0(m\phi)}r^4A_0^4}{\Delta_0^2}\\\nonumber
&\times \left\{\left(
\frac{a}{A_0}\right)'-\frac{L_0}{r^2}\frac{A_{0}^{\theta}}{A_0B_0^2}
\left(\frac{l}{L_0}+\frac{a^{\theta}}{A_{0}^{\theta}}-\frac{a}{A_0}
-\frac{2b}{B_0}\right)\right\}+\frac{\rho_{0(m\phi)}L_0^2A_0^2r^2}{\Delta_0^2}
\\\label{25}
&\times\left.\left(\frac{2l}{L_0}+\frac{2a}{A_0}-\frac{2\Delta_{p}}{\Delta_0}\right)
+\frac{\rho_{0(m\phi)}L_0^2A_0^2r^2}{\Delta_0^2}
\left\{\frac{l}{2}+\left(\frac{b}{B_0}\right)'\right\}\right]+\bar{E}_{1}=0.
\end{align}
The superscript $p$ indicates perturbed form, $\bar{E_1}$ represents
perturbed configurations of scalar field terms given in
Eq.(\ref{A10}).

\subsection{Spin-Independent Oscillations}

The local spinning of anisotropic system is calculated through the
vorticity tensor. For the axial symmetric spacetime with reflection,
the vorticity tensor can be expressed in terms of $k_{\mu}$ and
$\chi_{\mu}$ as
\begin{equation}\nonumber
\Omega_{\mu\nu}=\Omega(k_{\mu}\chi_{\nu}-\chi_{\mu}k_{\nu}),
\end{equation}
where
\begin{equation}\nonumber
\Omega=\frac{1}{2B\sqrt{\Delta}}\left(L'-2\frac{A'L}{A}\right),
\end{equation}
is the vorticity scalar. This shows that the spin-independent motion
occurs whenever $\Omega=0$ which is possible if $(L'
-2\frac{A'L}{A})=0$. This leads to
\begin{equation}\nonumber
\ln\left(\frac{L\tilde{K}}{A^{2}}\right)=0,
\end{equation}
where $\tilde{K}=\tilde{K}(t,\theta)$ is an arbitrary function of
integration. This implies that $L\tilde{K}=A^{2}$,
consequently, if $L=0$ we have $A=0$ and the existence of non-static
configuration of axial spacetime is disturbed. Therefore, for
$\Omega=0$, we take
\begin{equation}\nonumber
L\tilde{K}=A^{2},\quad L\neq0.
\end{equation}
Thus, oscillations convert into spin-independent form whenever,
$L\tilde{K}=A^{2}$ and $L\neq0$. This implies that spin-independent
oscillations depends upon reflection contribution. Equations
(\ref{23})-(\ref{25}) along with $L\tilde{K}=A^{2},~L\neq0$ provide
frequency, total density and equation of motion of spin-independent
oscillations of axisymmetric distribution.

\subsection{Stability Analysis}

Here, we discuss stability of oscillating collapsing axial system
(with reflection symmetry) in the presence of scalar field. We
assume that the system is perturbed adiabatically and satisfies the
equation of state \cite{15}
\begin{equation}\label{26}
\bar{p}=\Gamma\frac{p_0}{\rho_{0}+p_0}\bar{\rho},
\end{equation}
where the equation of state parameter ($\Gamma$) represents a
constant adiabatic index which calculates stiffness or rigidity in
the fluid. This equation with Eq.(\ref{24}) provides the perturbed
part of anisotropic stresses as follows
\begin{eqnarray}\nonumber
\bar{p}_{(m\phi)}&=&-\Gamma\frac{p_{0(m\phi)}}{\rho_{0(m\phi)}+p_{0(m\phi)}}
\left((F_{(m\phi)}+\bar{E}_{0(a)})iw+\bar{E}_{0(b)}\right)e^{iwt},\\\nonumber
\bar{\Pi}_{I(m\phi)}&=&-\Gamma\frac{\Pi_{I0(m\phi)}}{\rho_{0(m\phi)}+\Pi_{I0(m\phi)}}
\left((F_{(m\phi)}+\bar{E}_{0(a)})iw+\bar{E}_{0(b)}\right)e^{iwt},\\\nonumber
\bar{\Pi}_{II(m\phi)}&=&-\Gamma\frac{\Pi_{II0(m\phi)}}{\rho_{0(m\phi)}+\Pi_{II0(m\phi)}}
\left((F_{(m\phi)}+\bar{E}_{0(a)})iw+\bar{E}_{0(b)}\right)e^{iwt},\\\nonumber
\bar{\Pi}_{k\chi(m\phi)}&=&-\Gamma\frac{\Pi_{k\chi0(m\phi)}}{\rho_{0(m\phi)}+\Pi_{k\chi0(m\phi)}}
\left((F_{(m\phi)}+\bar{E}_{0(a)})iw+\bar{E}_{0(b)}\right)e^{iwt}.
\end{eqnarray}
Using these relations in the equation of motion
(\ref{25}), we obtain the collapse equation (hydrostatic equation)
of oscillating axial reflection system
\begin{equation}\label{27}
-\Gamma(\delta_{(m(BD))}(iw,r,\theta))
=-(\lambda_{(m(BD))}(iw,r,\theta)+\bar{E}_{1}).
\end{equation}
The quantity $\delta_{(m(BD))}$ shows pressure gradient forces and
anti-gravitational forces (due to matter as well as scalar field
distributions) coupled to adiabatic index whereas
$\lambda_{(m(BD))}+\bar{E}_{1}$ gives gravitational forces (forces
opposite to pressure gradients forces) mediated by matter as well as
scalar field contributions.  The values of these terms are given in
(\ref{A11}) and (\ref{A12}). The adiabatic index $\Gamma$ is taken
to be positive in order to balance the hydrostatic configurations
between pressure gradient as well as gravitational force.

\subsection*{Newtonian Approximations}

In order to evaluate stability criteria in the N limits, we use
approximation
$A_{0}=1,~B_{0}=1,~C_{0}=r,~L_{0}=r,~\Delta_{0}=r^{2},~\phi=\phi_{0}$
and $V(\phi)=V_{0}$. By using these limits in Eq.(\ref{27}), it
follows that
\begin{equation}\label{27'}
-\Gamma\left(\delta_{(m(BD))N}(iw,r,\theta)\right)=
-\left(\lambda_{(m(BD))N}(iw,r,\theta)+\bar{E}_{1(N)}\right),
\end{equation}
where the values of $\delta_{(m(BD))N},~\lambda_{(m(BD))N}$ and
$\bar{E}_{1(N)}$ are given in (\ref{A13})-(\ref{A15}). This provides
a hydrostatic condition which implies that the system collapses
whenever
\begin{equation}\nonumber
-\Gamma(\delta_{(m(BD))N})<-(\lambda_{(m(BD))N}+\bar{E}_{1(N)}),
\end{equation}
or
\begin{equation}\label{28}
\Gamma<\frac{-(\lambda_{(m(BD))N}+\bar{E}_{1(N)})}
{-\delta_{(m(BD))N}}.
\end{equation}
For $\Gamma>0$, we need to take
$|(\lambda_{(m(BD))N}+\bar{E}_{1(N)})|$ and $|\delta_{(m(BD))N}|$.
Thus the system remains unstable (collapses) as long as the
inequality (\ref{28}) holds. This implies that the instability
ranges in N limits can be calculated through the stiffness of fluid
(adiabatic index) which depends upon the configurations of pressure
gradient forces as well as anti-gravitational forces coupled to
adiabatic index and gravitational forces. These factors in turn
depend upon the energy density, anisotropies, reflection effects and
scalar field contributions. We can summarize the results as follows:
\begin{itemize}
\item If the gravitational forces $|(\lambda_{(m(BD))N}+\bar{E}_{1(N)})|$ are balanced by
anti-gravitational and pressure gradient forces
($\delta_{(m(BD))N}$) then (\ref{27'}) implies that $\Gamma=1$ and
the system is in complete hydrostatic equilibrium (remains stable).
\item If the anti-gravitational and pressure gradient distribution
related to stiffness parameter are greater than gravitational
contribution, then according to (\ref{27'}), the system becomes
unstable (but not collapses) for $0<\Gamma<1$.
\item Equation (\ref{27'}) implies that if
gravitational effects are greater than that of the
anti-gravitational and pressure gradient effects coupled to
adiabatic index, the system collapses leading to instability for
$\Gamma>1$.
\end{itemize}
In the case of spin-independent oscillations, the inequality
(\ref{28}) with $L\tilde{K}=A^{2}$ and $L\neq0$ provides the
criteria for an unstable spin-independent oscillations. The
numerical instability ranges ($0<\Gamma<1$ and $\Gamma>1$) remain
the same as calculated for spin-dependent oscillations.

\subsection*{Post-Newtonian Approximation}

In pN limits, we use approximations upto order of
$\frac{m_{0}}{r_{1}}$ (discarding terms having higher order of
$\frac{m_{0}}{r_{1}}$) as follows: $A_{0}=1-\frac{m_{0}}{r_{1}},~
B_{0}=1-\frac{m_{0}}{r_{1}},~\phi=\phi_{0}+\varphi$ and
$V=V_{0}+\varphi V'_{0}$ \cite{16}, $\varphi$ represents local
deviations of scalar field from $\phi_{0}$. The axial system becomes
unstable in pN limits if the adiabatic index satisfies the following
inequality
\begin{equation}\nonumber
\Gamma<\frac{(\lambda_{(m(BD))pN}+\bar{E}_{1(pN)})}
{\delta_{(m(BD))pN}},
\end{equation}
the values of $\delta_{(m(BD))pN}$ and
$(\lambda_{(m(BD))pN}+\bar{E}_{1(pN)})$ are given in
(\ref{A16})-(\ref{A18}). Similar to N case, the instability criteria
depends upon rigidity of the fluid and $\Gamma=1$ provides stable
configurations in pN regime while the system becomes unstable for
other values of adiabatic index.

\section{Concluding Remarks}

According to recent observation, DE controls the dynamics of
expansion of the present universe. General relativity is considered
as a fundamental theory for the description of various astrophysical
processes. It is an excellent theory of gravity which has gained
many achievements but said to break down at Planck length. Its
proposed DE candidate ``cosmological constant'' is not considered to
be compatible with the calculated vacuum energy of quantum fields.
This non-normalizable behavior of general relativity induces the
concept of alternative theories of gravity (alternative candidates
of DE) \cite{17}. These theories are constructed by incorporating
extra degrees of freedom in the Einstein-Hilbert Lagrangian density
either in geometrical (gravitational) or matter part. Some of these
DE models are Chaplygin gas, tachyon fields, quintessence, k-essence
and modified gravities such as $f(R)$ gravity, $f(T)$ theory,
Gauss-Bonnet gravity, $f(R,T)$ gravity and scalar-tensor theories.

Scalar-tensor theory of gravity is an alternative step to unify
theoretically gravity and quantum mechanics at high energies by
introducing scalar field as an extra degree of freedom in the
Einstein-Hilbert Lagrangian density. Brans-Dicke gravity is the
first proper scalar-tensor gravity assumed to be prototype of
alternative theory of Einstein gravity. The principal features of
this theory are the compatibility with Dirac's hypothesis and Mach's
principle, i.e., this theory believes on dynamical gravitational
coupling (dynamical gravitational constant) by means of dynamical
scalar field $\phi$ (extra force field) which allows distributions
of distant matter to affect the dynamics at a point. The basic idea
of this theory is that the inertial mass of an object is not an
intrinsic property of the object itself but is generated by the
gravitational effect of all the other matter in the universe. In
this way, this theory has generalized the Einstein gravity to
Machian one (compatible with Mach principle) and has provided
convenient solutions of many cosmic problems especially accelerated
expansion of the universe.

In general relativity, the effects of stellar rotations cannot be
neglected to fully investigate the formation of stars and black
holes. During evolution, the self-gravitating fluid passes through
many phases of dynamical activities that remain in hydrostatic
equilibrium for a short span. In this paper, we have studied
dynamical stability of non-static stellar model with axial
reflection symmetric anisotropic fluid distribution under the
influence of dynamical gravitational coupling through SBD gravity.
We have generalized the dynamical analysis of general relativity by
incorporating Mach's principle to explore effects of DE upon the
cosmic evolution

When a system is departed from its initial static phase, it becomes
perturbed and starts oscillating. We have explored oscillations of
the axial reflection configuration under time-frequency dependent
perturbation. It is shown that frequency of the rotating
oscillations depends upon anisotropic effects, reflection symmetric
and DE contribution represented by scalar field. The perturbed form
of conservation laws yields the total energy density and equation of
motion that depend upon the behavior of anisotropic effects as well
as frequency. We have also investigated spin-independent oscillation
and found that reflection configuration is the factor which controls
spin of the axial system.

In order to obtain viable models of the rotating system, we have
studied different instability ranges with the help of collapse
equation. It is found that the stable configurations of
spin-dependent oscillations depend upon the stiffness of the fluid
which in turn depends upon anisotropic effects, reflection parameter
as well as distribution of scalar field. The instability of
spin-independent oscillating system depends upon the rigidity of the
fluids due to anisotropy, reflection effects with constraints
$L\tilde{K}=A^{2},L\neq0$ as well as SBD gravity contribution. We
would like to mention here that dynamics of axial rotating system in
general relativity depends upon the anisotropic as well as
reflection effects but here the results are modified by the
inclusion of an extra field (scalar field) as a DE candidate. Thus
we can conclude that in the present accelerating universe, DE not
only controls the expansion among celestial objects but it also
affects stellar evolution.

According to Mach's principle, the dynamics of any evolving body in
the universe is not an intrinsic property, but the surrounding
distant matter also has its effect. In this way, it involves all the
surrounding stellar structures in the analysis. It would be
intersecting to explore collapsing phenomenon and its consequences
on the stellar objects according to Mach's principle to enhance the
study of stellar evolution in the presence of DE.

\section*{Appendix A}
\vspace{0.3cm}

\renewcommand{\theequation}{A\arabic{equation}}
\setcounter{equation}{0}

The values of scalar field energy terms $E_{0}(t,r,\theta),~
E_{1}(t,r,\theta)$ and $E_{2}(t,r,\theta)$ are given by
\begin{eqnarray}\nonumber
&&E_{0}(t,r,\theta)=\dot{v_{1}}+\dot{w_{1}}+x'_{1}+y'_{1}
+x^{\theta}_{3}+y^{\theta}_{3}+
\left(\frac{2B^{2}r^{2}A\dot{A}}{\Delta}+2\frac{L\dot{L}}{\Delta}
+\frac{\dot{B}}{B}+\frac{\dot{C}}{C}\right.\\\nonumber
&&\left.+\frac{LAA^{\theta}}{A}
+\frac{ABr^{2}\dot{B}}{\Delta}\right)(v_{1}+w_{1})+\left(3
\frac{B^{2}AA'r^{2}}{\Delta}+2\frac{A^{2}Br^{2}B'}{\Delta}
+\frac{2LL'}{\Delta}+\frac{B'}{B}\right.\\\nonumber
&&\left.+\frac{C'}{C}+\frac{A^{2}B^{2}r}{\Delta}\right)
(x_{1}+y_{1})+\left(\frac{3B^{2}AA^{\theta}r^{2}}{\Delta}
+\frac{3B\dot{B}r^{2}}{\Delta}+\frac{B^{\theta}}{B}
+\frac{LL^{\theta}}{\Delta}\right.\\\nonumber
&&\left.+\frac{C^{\theta}}{C}-\frac{LB\dot{B}r^{2}}{\Delta}
+\frac{A^{2}BB^{\theta}r^{2}}{\Delta}\right)(x_{3}+y_{3})+
(\frac{B^{3}r^{2}\dot{B}}{\Delta}-\frac{BLB^{\theta}}{\Delta})
(v_{2}+w_{2})\\\nonumber
&&+\left(\frac{B^{2}r^{2}L^{\theta}}{\Delta}
-\frac{B^{2}r^{2}\dot{B}}{\Delta} -\frac{LB^{\theta}}{\Delta}\right)
(v_{3}+w_{3}) +\left(\frac{CB^{2}r^{2}\dot{C}}{\Delta}
-\frac{LC^{\theta}}{\Delta}\right)\\\label{A1}
&&\times(v_{4}+w_{4})+\left(\frac{B^{2}r^{2}L'}{\Delta}
+\frac{LrB'}{\Delta}+\frac{LB}{\Delta}\right)(x_{2}+y_{2}),\\\nonumber
&&E_{1}(t,r,\theta)=(\dot{x}_{1}+\dot{y}_{1})+(v_{2}+w_{2})'
+(v_{1}+w_{1})^{\theta}+(v_{1}+w_{1})(\frac{AA'}{B^{2}})\\\nonumber
&&+\left(\frac{B^{2}r^{2}AA'}{\Delta}+\frac{3LL'}{2\Delta}+2\frac{B'}{B}+\frac{C'}{C}
+\frac{A^{2}BB'r^{2}}{\Delta}
+\frac{A^{2}B^{2}r}{\Delta}\right)(v_{2}+w_{2})\\\nonumber
&&-(v_{3}+w_{3}) \frac{r(rB'-B)}{B}+\left(\frac{3\dot{B}}{B}
+\frac{B^{2}r^{2}AA^{\theta}}{\Delta}+\frac{L\dot{L}}{\Delta}
-\frac{LAA^{\theta}}{\Delta}+\frac{AB\dot{B}r^{2}}{\Delta}\right)\\\nonumber
&&\times(x_{1}+y_{1})+\left(3\frac{B^{\theta}}{B}
+\frac{B^{2}Ar^{2}A^{\theta}}{\Delta}
+\frac{L\dot{B}}{\Delta}+\frac{LL^{\theta}}{\Delta}
-\frac{LB\dot{B}r^{2}}{\Delta}
+\frac{A^{2}BB^{\theta}r^{2}}{\Delta}\right)\\\label{A2}
&&\times(x_{2}+y_{2})-\frac{CC'}{B^{2}}(v_{4}+w_{4})
-\frac{L'}{B^{2}}(x_{3}+y_{3}),\\\nonumber
&&E_{2}(t,r,\theta)=(\dot{x}_{3}+\dot{y}_{3})+(x_{2}+y_{2})'
+(v_{3}+w_{3})^{\theta}+\left(\frac{AL'}{2\Delta}-
\frac{ALA'}{\Delta}\right)\\\nonumber &&\times(x_{1}+y_{1})
+\left(\frac{3AB\dot{B}r^{2}}{\Delta}
+\frac{B^{2}r^{2}A\dot{A}}{\Delta}+\frac{L\dot{L}}{\Delta}
+\frac{\dot{B}}{B}+\frac{\dot{C}}{C}
-\frac{2ALA^{\theta}}{\Delta}\right)\\\nonumber
&&\times(x_{3}+y_{3})+\left(2\frac{LL'}{\Delta}
+\frac{3A^{2}BB'r^{2}}{2\Delta}+\frac{3A^{2}B^{2}r}{\Delta}
+\frac{B'}{B}+\frac{C'}{C}\right)(x_{2}+y_{2})\\\nonumber
&&+\left(\frac{A^{2}BB^{\theta}r^{2}}{\Delta}
-\frac{3LL^{\theta}}{\Delta}-\frac{B^{2}r^{2}AA^{\theta}}{\Delta}
+\frac{Br^{2}L\dot{B}}{\Delta}
+\frac{B^{\theta}}{B}+\frac{C^{\theta}}{C}\right)(v_{3}+w_{3})\\\label{A3}
&&-\left(\frac{CL\dot{C}}{\Delta}
+\frac{A^{2}C^{\theta}}{\Delta}\right)(v_{4}+w_{4}).
\end{eqnarray}

In the perturbed configuration of $02$-component of Eq.(\ref{2}),
the resulting values of $l,~m$ and $n$ are
\begin{align}\label{A4}
&l=-\left[-\frac{r^2cB_0^2L_0^3}{C_0\Delta_0^2}-\frac{r^4cA_0^2B_0^4L_0}
{C_0\Delta_0^2}-\frac{L_0^3B_0r^2b}{\Delta_0^2}\right](x'_{3(p1)}+y'_{3(p1)}),
\\\nonumber
&m=i\left[-r^2aL_0^2A_0^2B_0^2
+\frac{L_0B_0A_0^2r^2B_{0}^{\theta}l}{\Delta_0^2} +\frac{C^\theta
L_0^4}{C_0\Delta_0^2}
+\frac{4L_0B_0^4r^2C_{0}^{\theta}l}{C_0\Delta_0^2}\right.\\\nonumber
&\left.+\frac{L_0A_0^2B_0^2r^2L_{0}^{\theta}c}{C_0A_0^2}
-\frac{r^2B_0^2aA_0C_{0}^{\theta}}{C_0\Delta_0^2}
+\frac{L_0A_0^2B_0r^2L_{0}^{\theta}b}{\Delta_0^2}-\frac{B_0^3b^\theta
A_0^2r^4}{\Delta_0^2}\right.\\\nonumber
&\left.+\frac{L_0^5C_{0}^{\theta}b}{B_0C_0\Delta_0^2}
+\frac{L_0^5B_{0}^{\theta}c}{B_0C_0\Delta_0^2} +\frac{b^\theta
L_0^4}{B_0\Delta_0^2} -\frac{r^4A_0^4B_0^5c^\theta }{\Delta_0^2}
+\frac{br^4A_0^2B_0^2B_{0}^{\theta}}{\Delta_0^2}
+\frac{br^4A_0^3B_0^3A_{0}^{\theta}}{\Delta_0^2}\right.\\\label{A5}
&\left.+\frac{cr^4B_0^4A_0^3A_{0}^{\theta}}{C_0\Delta_0^2}
+r^4bA_0^4B_0^3C_{0}^{\theta}
{C_0\Delta^2}\right](x_{3(p2)}+y_{3(p2)}),\\\nonumber
&n=-\frac{2r^4L_0A_0^3B_0A_0'B_0'}{\Delta_0^4}\left(\frac{l}{L_0}
+\frac{3a}{A_0}+\frac{b}{B_0}
+\frac{a'}{A_0}-\frac{2\Delta_p}{\Delta_0}\right)-\frac{3}{2}
\frac{r^2A_0^2L_0^2L_0'B_0'}{B_0\Delta_0^2}\\\nonumber
&\times\left(\frac{l'}{L_0'}+\frac{b'}{B_0'}+\frac{2a}{A_0}
+\frac{2l}{L_0}-\frac{2\Delta_p}{\Delta_0}
-\frac{b}{B_0}\right){\Delta_0^2}-\frac{br^2A_0^2L_0^2B_{0}^{\theta}}{\Delta_0^2}
-\frac{3r^2A_0A_0'L_0'L_0^2} {\Delta_0^2}\\\nonumber
&\times\left(\frac{a}{A_0}
+\frac{a'}{A_0'}+\frac{l'}{L_0'}+\frac{2l}{L_0}-\frac{2\Delta_p}{\Delta_0}\right)
+\frac{3r^2L_0L_0'^2A_0^2}{\Delta_0^2}\left(\frac{l}{L_0}+\frac{2l'}{L_0'}
+\frac{2a}{A_0}-\frac{2\Delta_p}{\Delta_0}\right)\\\nonumber
&+\frac{3rA_0^2B_0'L_0^3}{B_0\Delta_0^2}\left(\frac{2a}{A_0}
+\frac{b'}{B_0'}-\frac{b}{B_0}+3l\Delta_{p}L_{0}\Delta_{0}\right)
-\frac{A_0^2L_0^2B_{0}^{\theta}L_{0}^{\theta}T}{B-0\Delta_0^2}\left(
\frac{2a}{A_0}+\frac{2l}{L_0}
+\frac{b^\theta}{B_{0}^{\theta}}\right.\\\nonumber
&\left.+\frac{l^{\theta}}{L_{0}^{\theta}}
-\frac{b^2\Delta_p}{B_0\Delta_0}\right)
-\frac{r^2B_0'C_0'A_0^2L_0^3}{B_0C_0\Delta_0^2}\left(\frac{b'}{B_0'}
+\frac{c'}{C_0'}+\frac{2a}{A_0}+\frac{3l}{L_0}
-\frac{b}{B_0}-\frac{c}{C_0}-\frac{2\Delta_p}{\Delta_0}\right)\\\nonumber
&-\frac{L_0A_0^4r^4B_0'^2}{\Delta_0^4}\left(\frac{l}{L_0}+\frac{4a}{A_0}
+\frac{2b}{B_0}-\frac{2\Delta_p}{\Delta_0}\right)
+\frac{r^4L_0B_0^2A_0^3A_0'C_0'}{C_0\Delta_0^2}\left(\frac{l}{L_0}
+\frac{2b}{B_0}+\frac{3a}{A_0}\right.\\\nonumber
&\left.+\frac{a'}{A_0'}
+\frac{c'}{C_0'}-\frac{c}{C_0}-\frac{2\Delta_p}{\Delta_0}\right)
+\frac{r^2A_0A_0'C_0'L_0^3}{C_0\Delta_0^2}
\left(\frac{3l}{L_0}+\frac{a'}{A_0'}+\frac{a}{A_0}
+\frac{c'}{C_0'}-\frac{c}{C_0}\right.\\\nonumber
&\left.-\frac{2\Delta_p}{\Delta_0}\right)
+\frac{r^2A_0A_0'C_0'L_0^3}{C_0\Delta_0^2}\left(
\frac{2a'}{A_0'}+\frac{3l}{L_0}-\frac{2\Delta_p}{\Delta_0}\right)
+\frac{A_0^2B_{0}^{\theta}C_{0}^{\theta}L_0^3}{B_0C_0\Delta_0^2}
\left(\frac{2a}{A_0}+\frac{b^\theta}{B_{0}^{\theta}}\right.\\\nonumber
&\left. +\frac{c^\theta}{C_{0}^{\theta}}-\frac{b}{B_0}
-\frac{c}{C_0}+\frac{3l}{L_0}-\frac{2\Delta_p}{\Delta_0}\right)
-\frac{4A_0^2L_0^2L_{0}^{\theta}C_{0}^{\theta}}{C_0\Delta_0^2}
\left(\frac{2a}{A_0}+\frac{2l}{L_0}
+\frac{l^{\theta}}{L_{0}^{\theta}}+\frac{c^\theta}{C_{0}^{\theta}}\right.\\\nonumber
&\left.-\frac{c}{C_0}-\frac{2\Delta_p}{\Delta_0}\right)+\frac{rC_0'L_0^3}{C_0\Delta_0^2}
\left(\frac{2a}{A_0}+\frac{c'}{C_0'}-\frac{c}{C_0}
+\frac{3l}{L_0}-\frac{2\Delta_p}{\Delta_0}\right)
+\frac{L_0A_0^4B_0^2C_0'r^3}{C_0\Delta_0^2}\\\nonumber &\times\left(
\frac{l}{L_0}+\frac{4a}{A_0}+\frac{2b}{B_0}+\frac{c'}{C_0'}
-\frac{c}{C_0}-\frac{2\Delta_p}{\Delta_0}\right)
-\frac{B_0'C_0'L_0^5}{B_0^3C_0\Delta_0^2}
\left(\frac{b'}{B_0'}+\frac{c'}{C_0'}+\frac{5l}{L_0}\right.\\\nonumber
&\left.-\frac{3b}{B_0}
-\frac{c}{C_0}-\frac{2\Delta_p}{\Delta_0}\right)+\frac{A_0^2L_0^3}{\Delta_0^2}
\left(\frac{2a}{A_0}+\frac{3l}{L_0}
-\frac{2\Delta_p}{\Delta_0}\right)-\frac{L_0^3L_0'}{B_0^2\Delta_0^2}
\left(\frac{3l}{L_0}\right.\\\nonumber &\left.+\frac{2l'}{L_0'}
-\frac{2\Delta_p}{\Delta_0}-\frac{2b}{B_0}\right)
-\frac{br^2B_0A_0^2L_0^2C_{0}^{\theta}}{C_0\Delta_0^2}
+\frac{r^4A_0^3B_0^2A_0'L_0'}{2L_0^2}\left(\frac{2b}{B_0}
+\frac{3a}{A_0}\right.\\\nonumber
&\left.+\frac{a'}{A_0'}+\frac{l}{L_0'}
-\frac{2\Delta_p}{\Delta_0}\right)+\frac{4r^4B_0A_0^4B_0'L_0'}
{\Delta_0^2}\left(\frac{3l}{L_0}+\frac{a}{A_0}
+\frac{a^\theta}{A_{0}^{\theta}}+\frac{b^\theta}{B_{0}^{\theta}}
-\frac{b}{B_0}\right)\\\nonumber
&-\frac{3rA_0^2L_0^2L_0'}{2\Delta_0^2}
\left(\frac{2a}{A_0}+\frac{l'}{L_0'}+\frac{2l}{L_0}
-\frac{2\Delta_p}{\Delta_0}\right)+\frac{2r^3A_0^4B_0^2L_0'}{\Delta_0^2}\\\nonumber
& \times\left(\frac{4a}{A_0}+\frac{2b}{B_0}+\frac{l'}{L_0'}
-\frac{2\Delta_p}{\Delta_0}\right)
+\frac{L_0^3A_0A_{0}^{\theta}C_{0}^{\theta}}{C_0\Delta_0^2}
\left(\frac{3l}{L_0}+\frac{a}{A_0}+\frac{a^\theta}{A_{0}^{\theta}}
+\frac{c^\theta}{C_{0}^{\theta}}\right.\\\nonumber
&\left.-\frac{c}{C_0}-\frac{2\Delta_p}{\Delta_0}\right)
-\frac{r^4A_0^4B_0^2L_0'C_0'T}{C_0\Delta_0^2}\left(
\frac{l'}{L_0'}+\frac{c'}{C_0'}+\frac{4a}{A_0}+\frac{2b}{B_0}-\frac{c}{C_0}
-\frac{2\Delta_p}{\Delta_0}\right)\\\nonumber
&-\frac{L_0'B_0'L_0^4}{2B_0^3\Delta_0^2}
\left(\frac{l'}{L_0'}+\frac{b'}{B_0'}+\frac{4l}{L_0}
-\frac{3b}{B_0}-\frac{2\Delta_p}{\Delta_0}\right)
+\frac{2L_0'C_0'L_0^4}{B_0^2C_0\Delta_0^2}\left(\frac{l'}{L_0'}
+\frac{c'}{C_0'}\right.\\\nonumber
&\left.+\frac{4l}{L_0}-\frac{2b}{B_0}
-\frac{c}{C_0}-\frac{2\Delta_p}{\Delta_0}\right)
-\frac{r^4A_0^4B_0^2L_0''}{2\Delta_0^2}
\left(\frac{c}{C_0}+\frac{l''}{L_0''}+\frac{2b}{B_0}
+\frac{4a}{A_0}-\frac{2\Delta_p}{\Delta_0}\right)\\\nonumber &
+\frac{L_0''L_0^4}{2\Delta_0^2B_0^2}\left(\frac{l''}{L_0''}
+\frac{4l}{L_0}-\frac{2\Delta_p}{\Delta_0}-\frac{2b}{B_0}\right)
+\frac{C_0''L_0^5}{\Delta_0^2B_0^2C_0}\left(\frac{c''}{C_0''}+\frac{5l}{L_0}-\frac{2b}{B_0}
-\frac{c}{C_0}\right.\\\nonumber
&\left.-\frac{2\Delta_p}{\Delta_0}\right)
+\frac{A_0^2L_0^3C_{0}^{\theta\theta}}{C_0\Delta_0^2}\left(\frac{2a}{A_0}
+\frac{c^{\theta\theta}}{C_{0}^{\theta\theta}}
+\frac{3l}{L_0}-\frac{2\Delta_p}{\Delta_0}-\frac{c}{C_0}\right)
+\frac{r^2A_0A_0''L_0^3}{\Delta_0^2}
\\\nonumber
&\times\left(\frac{a}{A_0}
+\frac{a''}{A_0''}+\frac{3l}{L_0}-\frac{\Delta_p}{\Delta_0}\right)
+\frac{r^2A_0^2B_0''L_0^3}{B_0\Delta_0^2}
\left(\frac{2a}{A_0}+\frac{b''}{B_0''}+\frac{3l}{L_0}-\frac{b}{B_0}
-\frac{\Delta_p}{\Delta_0}\right)\\\nonumber
&+\frac{r^4A_0^3B_0^2L_0A_0''}{\Delta_0^2}\left(\frac{l}{L_0}
+\frac{2b}{B_0}+\frac{3a}{A_0}+\frac{a''}{A_0''}
-\frac{\Delta_p}{\Delta_0}\right)
+\frac{r^2A_0^4B_0B_{0}^{\theta}}{\Delta_0^2}
\left(\frac{l}{L_0}\right.\\\nonumber
&\left.+\frac{4a}{A_0}+\frac{b}{B_0}+\frac{b^\theta}{B_{0}^{\theta}}
-\frac{\Delta_p}{\Delta_0}\right)+\frac{L_0A_0^4B_0r^4B_0''}{\Delta_0^2}
\left(\frac{l}{L_0}+\frac{4a}{A_0}-\frac{b}{B_0}+\frac{b''}{B_0''}
-\frac{\Delta_p}{\Delta_0}\right)\\\nonumber
&+\frac{A_0^2L_0^3L_{0}^{\theta\theta}}{B_0\Delta_0^2}\left(\frac{2a}{A_0}
+\frac{b^{\theta\theta}}{L_{0}^{\theta\theta}}
+\frac{3l}{L_0}-\frac{b}{B_0}-\frac{\Delta_p}{\Delta_0}\right)
+\frac{2C_0''A_0^2r^2L_0^3}{C_0\Delta_0^3}\left(
\frac{c''}{C_0''}-\frac{\Delta_p}{\Delta_0}\right.\\\nonumber&\left.
-\frac{c}{C_0}+\frac{2a}{A_0}+\frac{3l}{L_0}\right)
+\frac{L_0C_0''B_0^2A_0^4r^4}{C_0\Delta_0^2}\left(\frac{l}{L_0}
+\frac{c''}{C_0''}+\frac{2b}{B_0} +\frac{4a}{A_0}-\frac{c}{C_0}
-\frac{2\Delta_p}{\Delta_0}\right)\\\nonumber &
+\frac{4L_0A_0^4B_0^2r^2C_{0}^{\theta\theta}}{C_0\Delta_0^2}
\left(\frac{l}{L_0}+\frac{4a}{A_0}+\frac{2b}{B_0}
+\frac{c^{\theta\theta}}{C_{0}^{\theta\theta}}-\frac{c}{C_0}
-\frac{\Delta_p}{\Delta_0}\right)
=-\left[\rho_{0(m\phi)}\right.\\\nonumber&\left.\times\left\{\frac{b}{B_0}
+\frac{c}{C_0}+\frac{1}{\Delta_0}\left(r^2aA_0^2B_0^2+lL_0+r^2bB_0L_0\right)\right\}
+(\rho_{0(m\phi)}+p_{0(m\phi)})\right.\\\nonumber
&\times\frac{A_0^2B_0^2}{\Delta_0^2}\left\{r^2\left(\frac{2b}{B_0}
+\frac{2c}{C_0}\right)+\frac{A_0^2}{A_0^2B_0^2}
\left(\frac{b}{B_0}+\frac{l}{L_0}-\frac{a}{A_0}
+\frac{c}{C_0}\right)\right\}+\frac{\Pi_{I0(m\phi)}}{3}\\\nonumber
&\times\left.\left(\frac{b}{B_0}-\frac{c}{C_0}\right)
+\frac{\Pi_{II0(m\phi)}}{3\Delta_0}\left\{r^2A_0^2B_0^2\left(\frac{b}{B_0}
-\frac{c}{C_0}\right)+L_0^2\left(\frac{l}{L_0}-\frac{a}{A_0}
-\frac{c}{C_0}\right)\right\}\right]\\\label{A6}
&+(x_{3(p3)}+y_{3(p3)}).
\end{align}
Here the scalar field stress with subscript $p1,~p2$ indicates
perturbed values coupled to $w^{2},~w$, respectively and $p3$ shows
otherwise situation. The values of $F_{(m\phi)},~\bar{E}_{0(a)}$ and
$\bar{E}_{0(b)}$ given in Eq.(\ref{24}) are as follows
\begin{eqnarray}\nonumber
&&F_{(m\phi)}=-\left[\rho_{0(m\phi)}\left\{\frac{b}{B_0}+\frac{c}{C_0}+\frac{1}{\Delta_0}
\left(r^2aA_0B_0^2+l
L_0+r^2bB_0A_0^2\right)\right\}\right.\\\nonumber&&+
(\rho_{0(m\phi)}+p_{0(m\phi)})\frac{A_0^2B_0^2}{L_0}
\left\{r^2\left(\frac{2b}{B_0}+\frac{2c}{C_0}\right)+\frac{L_0^2}{A_0^2B_0^2}
\left(\frac{b}{B_0}+\frac{l}{L_0}-\frac{a}{A_0}\right.\right.\\\nonumber
&&\left.\left.+\frac{c}{C_0}\right)\right\}
+\frac{\Pi_{I0(m\phi)}}{3}\left(\frac{b}{B_0}-\frac{c}{C_0}\right)
+\frac{\Pi_{II0(m\phi)}}{3L_0}\left\{r^2A_0^2B_0^2\left(\frac{b}{B_0}
-\frac{c}{C_0}\right)\right.\\\label{A7}
&&\left.\left.+L_0^2\left(\frac{l}{L_0}-\frac{a}{A_0}
-\frac{c}{C_0}\right)\right\}\right],\\\nonumber
&&\bar{E}_{0(a)}=(\dot{v}_{1(ap)}+\dot{w}_{1(ap)})+(x'_{1(ap)}+y'_{1(ap)})
+(x^{\theta}_{3(ap)}+y^{\theta}_{3(ap)})+\left(\frac{c}{C_{0}}
\right.\\\nonumber &&\left.+\frac{b}{B_{0}}
+\frac{2A_{0}B_{0}ar^{2}}{\Delta_{0}}+\frac{B_{0}A^{2}_{0}b}{\Delta_{0}}
+\frac{2lL_{0}}{\Delta_{0}}\right)(v_{1(a0)}+w_{1(a0)})+\left(\frac{C'_{0}}{C_{0}}
+\frac{B'_{0}}{B_{0}}\right.\\\nonumber
&&\left.+\frac{2B^{3}_{0}r^{2}A_{0}A'_{0}}{\Delta_{0}}
+\frac{2L_{0}L'_{0}}{\Delta_{0}}\right)(x_{1(a0)}+y_{1(a0)})
+\frac{bB^{2}_{0}L_{0}r^{2}}{\Delta_{0}}
+\left(\frac{C^{\theta}_{0}}{C_{0}}
+\frac{B^{\theta}_{0}}{B_{0}}\right.\\\nonumber &&\left.
+\frac{\Delta^{\theta}_{0}}{\Delta_{0}}\right)(v_{(20)}
+w_{(20)})(\frac{B^{2}_{0}r^{2}}{\Delta_{0}})+
(x_{3(ap)}+y_{3(ap)})+(v_{2(ap)}+w_{2(ap)})\\\nonumber &&\times
(\frac{-2L_{0}B_{0}}{\Delta_{0}})+\left[\frac{-2B^{2}_{0}r^{2}L^{\theta}_{0}}{\Delta_{0}}
+\frac{L^{2}_{0}B_{0}r^{2}B^{\theta}_{0}}{\Delta_{0}}\right]
(x_{3(ap)}+y_{3(ap)})+(x_{3(a0)}\\\nonumber && +y_{3(a0)})
(\frac{B^{2}_{0}r^{4}b}{\Delta_{0}})+(x_{4(a0)}+y_{4(a0)})(\frac{C_{0}B^{2}_{0}r^{2}c}{\Delta_{0}})
-\frac{2L_{0}C^{\theta}_{0}}{\Delta_{0}}(x_{4(ap)}\\\label{A8}
&&+y_{4(ap)})
+(x_{(2ap)}+y_{(2ap)})\left[\frac{-B^{2}_{0}r^{2}L'_{0}}{\Delta_{0}}
+\frac{L_{0}r^{2}B_{0}B'_{0}}{\Delta_{0}}+\frac{L_{0}B^{2}_{0}r^{2}}{
\Delta_{0}}\right],\\\nonumber
&&\bar{E}_{0(b)}=\left[\frac{C^{\theta}_{0}}{C_{0}}+\frac{B^{\theta}_{0}}{B_{0}}
+\frac{\Delta^{\theta}_{0}}{\Delta_{0}}
+\frac{2B^{2}_{0}r^{2}A_{0}A'_{0}}{\Delta_{0}}\right]
(x_{3(bp)}+y_{3(bp)})\\\nonumber &&+(v_{2(bp)}+w_{2(bp)})
\left[\frac{-2L_{0}B^{\theta}_{0}}{\Delta_{0}}\right]
+\left[\frac{-2B^{2}_{0}r^{2}L^{\theta}_{0}}{\Delta_{0}}
+\frac{L_{0}B_{0}r^{2}L_{0}B^{\phi}_{0}}{\Delta_{0}}\right]
(x_{3(bp)}\\\nonumber&&+y_{3(bp)})+(v_{2(b0)}+w_{2(b0)})
\left[\frac{l}{L_{0}}+\frac{b}{B_{0}}
+\frac{\Delta'_{0}}{\Delta_{0}}\right]
-\frac{2L_{0}C^{\theta}_{0}}{\Delta_{0}}(v_{(4bp)}+w_{(4bp)})\\\nonumber
&&+(x_{2(bp)}+y_{2(bp)})\left(-B^{2}_{0}r'L'_{0}
+\frac{r^{2}B^{2}_{0}B'_{0}}{\Delta_{0}}
+\frac{L_{0}B^{2}_{0}r^{2}}{\Delta_{0}}\right)+(x_{2(0)} +y_{2(0)})
\left(-B^{2}_{0}\right.\\\nonumber &&\times\left.
r^{2}L'_{0}\left(\frac{b^{\theta}}{B^{\theta}}
+\frac{l'}{L'_{0}}\right)
+L_{0}r^{3}B_{0}B'_{0}\left(\frac{b'}{B'_{0}}+\frac{l}{L_{0}}\right)
+L_{0}B^{2}_{0}r^{2}\left(\frac{l}{L_{0}}+\frac{2b}{B_{0}}\right)\right)\\\label{A8}
&&+(x^{\theta}_{3(bp)}+y^{\theta}_{3(bp)}).
\end{eqnarray}
The subscript $a0$ and $ap$ denote scalar field unperturbed as well
as perturbed terms coupled to frequency while $b0$ and $bp$ show
otherwise scalar field static and perturbed configurations. The
value of scalar field term $\bar{E}_{1}$ is given by
\begin{eqnarray}\nonumber
&&\bar{E}_{1}=(\dot{x}_{(1p)}+\dot{y}_{(1p)})+(v'_{1(p)}+w'_{1(p)})
+(x_{2(p)}+y_{2(p)})+(v_{1(0)}+w_{1(0)})^{\theta}
\frac{A_{0}A'_{0}}{B^{2}_{0}}\\\nonumber &&\times
e^{iwt}\left(2\frac{b}{B_{0}}+\frac{a}{A_{0}}
+\frac{a'}{A'_{0}}\right) +(v_{1(p)}+w_{1(p)})
\frac{A_{0}A'_{0}}{B^{2}_{0}}+(v_{2(0)}+w_{2(0)})e^{iwt}\\\nonumber
&&\times
\left[\frac{2b'}{B_{0}B'_{0}}-\frac{B'_{0}b}{B^{2}_{0}}\right]
+(v_{2(0)}+w_{2(0)})e^{iwt}
\left[\frac{B^{2}_{0}r^{2}A_{0}A'_{0}}{\Delta_{0}}
\left(\frac{2b}{B^{2}_{0}}+\frac{a}{A_{0}}+\frac{a'}{A'_{0}}
-\frac{\Delta_{p}}{\Delta_{0}}\right)\right.\\\nonumber
&&\left.+\frac{2L_{0}L'_{0}}{\Delta_{0}}\left(\frac{l}{L_{0}}
+\frac{l'}{L'_{0}} +\frac{\Delta_{p}}{\Delta_{0}}\right)
+\frac{A^{2}_{0}B_{0}B'_{0}r^{2}}{\Delta_{0}}\left(2\frac{a}{A_{0}}
+\frac{b}{B_{0}}+\frac{b'}{B'_{0}}
-\frac{\Delta_{p}}{\Delta_{0}}\right)
+\frac{c^{\theta}}{C_{0}}\right.\\\nonumber &&\left.
+\frac{A^{2}_{0}B^{2}_{0}r}{\Delta_{0}}
\left[\frac{2a}{A_{0}}+2\frac{b}{B_{0}}
-\frac{\Delta_{p}}{\Delta_{0}}\right]\right]
+(v_{2(p)}+w_{2(p)})\left[2\frac{B'_{0}}{B_{0}}
+\frac{B^{2}_{0}r^{2}A_{0}A'_{0}}{\Delta_{0}}
+\frac{2L_{0}L'_{0}}{\Delta_{0}}\right.\\\nonumber
&&\left.+\frac{A^{2}_{0}B_{0}r^{2}B'_{0}}{\Delta_{0}}\right]
+(v_{3(p)}+w_{3(p)})\left(-r^{2}\frac{B'_{0}}{B_{0}}-r\right)
-(v_{4(p)}+w_{2(p)})\frac{C_{0}C'_{0}}{B^{2}_{0}}\\\nonumber
&&+(v_{3(0)}+w_{3(0)})e^{iwt}
\left(\frac{rb'}{B_{0}}-\frac{b}{B}_{0}\right)-(v_{4(p)}+w_{4(p)})
\left[\frac{C'_{0}c}{C_{0}}+\frac{c'}{C_{0}}-\frac{2b}{B_{0}}\right]\\\nonumber
&&+\left(\frac{-L_{0}A_{0}A^{\theta}_{0}}{\Delta_{0}}
+\frac{A_{0}B_{0}r^{2}B^{\theta}_{0}}{\Delta_{0}}\right)
(x_{1(p)}+y_{1(p)})+e^{iwt}\left[3\frac{L'_{0}}{B^{2}_{0}}
\left(\frac{l'}{L'_{0}}-\frac{2b}{B_{0}}\right)\right.\\\nonumber
&&\left.+\frac{B'_{0}}{B_{0}}\left(\frac{b'}{B'_{0}}
-\frac{b}{B_{0}}\right) +\frac{L_{0}L'_{0}}{\Delta_{0}}
\left(\frac{l}{L_{0}}+\frac{l'}{L'_{0}}
-\frac{\Delta_{p}}{\Delta_{0}}\right)
+\frac{A^{2}_{0}B_{0}r^{2}B'_{0}}{\Delta_{0}}\left(\frac{2a}{A_{0}}
\right.\right.\\\nonumber &&\left.\left.
+\frac{b}{B_{0}}+\frac{b'}{B'_{0}}-\frac{\Delta_{p}}{\Delta_{0}}\right)
+\frac{A^{2}_{0}B^{2}_{0}r}{\Delta_{0}}
\left[\frac{2a}{A_{0}}+\frac{2b}{B_{0}}\right]\right](x_{2(0)}+y_{2(0)})
+(x_{2(p)}\\\label{A10} && +y_{2(p)})
\left[3\frac{L'_{0}}{B^{2}_{0}}
+\frac{B'_{0}}{B_{0}}+\frac{L_{0}L'_{0}}{\Delta_{0}}
+\frac{A^{2}_{0}B_{0}r^{2}B'_{0}}{\Delta_{0}}
+\frac{A^{2}_{0}B^{2}_{0}r}{\Delta_{0}}\right].
\end{eqnarray}
The values of $\delta_{(m(BD))}$ and $\lambda_{(m(BD))}$ are
\begin{eqnarray}\nonumber
&&\delta_{(m(BD))}=\frac{1}{B_0^2}\left(\frac{p_{0(m\phi)}\beta}{\rho_{0(m\phi)}+p_{0(m\phi)}}
+\frac{4\Pi_{I0(m\phi)}\beta}{9(\rho_{0(m\phi)}+\Pi_{I0(m\phi)})}\right.\\\nonumber
&&\left.+\frac{2\Pi_{II0(m\phi)}\beta}{9(\rho_{0(m\phi)}+\Pi_{II0(m\phi)})}\right)'
-\left[\frac{p_0}{\rho_{0(m\phi)}+p_{0(m\phi)}}+\frac{4\Pi_{I0(m\phi)}}
{9(\rho_{0(m\phi)}+\Pi_{I0(m\phi)})}\right.\\\nonumber
&&\left.+\frac{2\Pi_{II0(m\phi)}}{9(\rho_{0(m\phi)}+\Pi_{II0(m\phi)})}\right]\frac{\beta
}{B_0^2}\left\{\frac{C_0'}{C_0}
+\frac{3L_0L_0'}{2\Delta_0}\right.\left.+\frac{r^2A_0^2B_0^2}{\Delta_0}\left(\frac{A_0'}{A_0}
+\frac{1}{r}\right)\right\}\\\nonumber
&&+\frac{r^2A_0B_0^3}{\Delta^{\frac{3}{2}}_0}
\left(\frac{\Pi_{k\chi0(m\phi)}\beta}{\rho_{0(m\phi)}+\Pi_{k\chi0(m\phi)}}\right)^\theta
+\frac{\Pi_{II0(m\phi)}\beta}{\rho_{0(m\phi)}+\Pi_{II0(m\phi)}}
\frac{r^2A_0B_0^3}{\Delta_0^{\frac{3}{2}}}\\\label{A11}
&&\times\left\{\frac{A_{0}^{\theta}}{A_0}
+\frac{6B_{0}^{\theta}}{B_0}\right.+\frac{C_{0}^{\theta}}{C_0}
+\frac{4L_0L_{0}^{\theta}}{\Delta_0}
\left.+\frac{4r^2A_0B^2_0}{\Delta_0}\left(\frac{A_{0}^{\theta}}{A_0}
+\frac{B_{0}^{\theta}}{B_0}\right)\right\},\\\nonumber
&&\lambda_{mBD}=\left[\frac{\beta
r^4A_0^4}{\Delta_0^2}\left(\frac{A_0'}{A_0}-\frac{L_0A_{0}^{\theta}}
{r^2A_0B_0^2}\right)\right.\frac{2b}{B_0^3}
\left\{{p_0}'+\frac{2}{9}(2{\Pi}_{I0(m\phi)}'+{\Pi}_{II0(m\phi)}')\right\}\\\nonumber
&&-\left[\left\{\left(\frac{c}{C_0}\right)'+\frac{3L_0L_0'}{2\Delta_0}
\left(\frac{l}{L_0}+\frac{l'}{L_0'}
-\frac{\Delta_{p}}{\Delta_0}\right)
\right\}\right.+\frac{r^2A_0^2B_0^2}{\Delta_0^2}
\left(\frac{2a}{A_0}+\frac{2b}{B_0}\right.\\\nonumber &&\left.
-\frac{\Delta_{p}}{\Delta_0}\right)\left(\frac{A_0'}{A_0}+\frac{1}{r}
\right)+\frac{r^2A_0^2B_0^2}{\Delta_0}\left.\left(\frac{a}{A_0}
+\frac{b}{B_0}\right)'\right]\frac{1}{B_0^2}
\left\{{p_{0(m\phi)}}+\frac{2}{9}(2{\Pi}_{I0(m\phi)}\right.\\\nonumber
&&\left.+{\Pi}_{II0(m\phi)}) \right\}-\frac{2b}{B_0^2}
\left\{{p_{0(m\phi)}}+\frac{2}{9}(2{\Pi}_{I0(m\phi)}+{\Pi}_{II0(m\phi)})\right\}
\left\{\frac{C_0'}{C_0}+\frac{3L_0L_0'}{2\Delta_0}\right.\\\nonumber
&&\left.+\frac{r^2A_0^2B_0^2}{\Delta_0}\left(\frac{A_0'}{A_0}\right.\right.
\left.\left.+\frac{1}{r}
\right)\right\}+\frac{r^2A_0B_0^3}{\Delta^{\frac{3}{2}}_0}{\Pi}_{k\chi0(m\phi)}^{\theta}
\left(\frac{a}{A_0}+\frac{3b}{B_0}-\frac{3\Delta_{p}}{\Delta_0}\right)\\\nonumber
&&+\frac{r^2A_0B_0^3\Pi_{k\chi0(m\phi)}}{\Delta_0^{\frac{3}{2}}}
\left(\frac{a}{A_0}+\frac{3b}{B_0}
-\frac{3\Delta_{p}}{\Delta_0}\right)\left\{\frac{A_{0}^{\theta}}{A_0}
+\frac{6B_{0}^{\theta}}{B_0}+\frac{C_{0}^{\theta}}{C_0}
+\frac{4L_0L_{_0}^{\theta}}{\Delta_0}\right.\\\nonumber
&&\left.+\frac{4r^2A^2_0B^2_0}{\Delta_0}
\left(\frac{A_{0}^{\theta}}{A_0}+\frac{B_{0}^{\theta}}{B_0}\right)\right\}
+\frac{r^2A_0B_0^3}{\Delta^{\frac{3}{2}}_0}{\Pi}_{k\chi0(m\phi)}
\left[\frac{6B_{0\theta}}{B_0}\left(\frac{b_\theta}{B_{0\theta}}
+\frac{b}{B_0}\right)\right.\\\nonumber
&&\times\left.\left(\frac{a}{A_0}
+\frac{c}{C_0}\right)^\theta+\frac{4L_0L_{0\theta}}{\Delta_0}
\left(\frac{l}{L_0}+\frac{l^\theta}{L_{0}^{\theta}}
-\frac{\Delta_{p}}{\Delta_0}\right)+\frac{4r^2A_0^2B_0^2}{\Delta_0}
\left(\frac{2a}{A_0}+\frac{2b}{B_0}\right.\right.\\\nonumber
&&\left.\left. -\frac{\Delta_{p}}{\Delta_0}\right)
\left(\frac{a}{A_0}+\frac{b}{B_0}\right)^\theta\right]
-\frac{\rho_{0(m\phi)}r^4A_0^4}{\Delta_0^2}\left\{\left(
\frac{a}{A_0}\right)'-\frac{L_0}{r^2}\frac{A_{0}^{\theta}}{A_0B_0^2}
\left(\frac{l}{L_0}+\frac{a^\theta}{A_{0}^{\theta}}\right.\right.\\\nonumber
&&\left.\left.-\frac{a}{A_0}
-\frac{2b}{B_0}\right)\right\}\left.+\frac{\rho_{0(m\phi)}L_0^2A_0^2r^2}{\Delta_0^2}
\left(\frac{2l}{L_0}+\frac{2a}{A_0}-\frac{2\Delta_{p}}{\Delta_0}\right)
+\frac{\rho_{0(m\phi)}L_0^2A_0^2r^2}{\Delta_0^2}\right.\\\label{A12}
&&\left.\times\left(\frac{l}{L_0}+\frac{b}{B_0}\right)'\right.
-L_0^2r^2\frac{A_0^2}{\Delta_0^2}\beta
\left.\left\{\frac{L_0'}{2L_0}+\frac{1}{r}+\frac{B_0'}{B_0}\right\}\right].
\end{eqnarray}
Here $\beta=\left[(F_{(m\phi)}
+\bar{E}_{0(a)})iw+\bar{E}_{0(b)}\right]$.

The values of $\delta_{(m(BD))N},~\lambda_{(m(BD))N}$ and
$\bar{E}_{1(N)}$ are
\begin{eqnarray}\nonumber
&&\delta_{(m(BD))N}=\left[\beta_{(N)}\left\{{p_{0(m\phi)}}+\frac{2}{9}
(2{\Pi}_{I0(m\phi)}+{\Pi}_{II0(m\phi)})\right\}\right]'
+\frac{11\beta_{(N)}}{4r}\left\{{p_{0(m\phi)}}\right.\\\label{A13}
&&\left.+\frac{2}{9} (2{\Pi}_{I0(m\phi)}+{\Pi}_{II0(m\phi)})\right\}
-\left[\frac{\Pi_{k\chi0(m\phi)}}{2\sqrt{2}r}\beta_{(N)}\right]^\theta,\\\nonumber
&&\lambda_{(m(BD))N}=\left[ \left\{{p_{0(m\phi)}}+\frac{2}{9}
(2{\Pi}_{I0(m\phi)}+{\Pi}_{II0(m\phi)})\right\}\left\{\frac{1}{2}(a+b)'
\right.\right.\\\nonumber
&&\left.\left.+\left(\frac{c}{r}\right)'+\frac{1}{2r}
\left(2a+11b-\frac{\Delta_{p}}{2r^2}\right)\right\}
+\frac{\Pi_{k\chi0(m\phi)}}{2\sqrt{2}}\left[2(a+b)^\theta\left(2a+2b\right.\right.\right.\\\nonumber
&&\left.\left.\left. -\frac{\Delta_{p}}{2r^2}\right)\right]
\frac{3}{4r}\left(l'+\frac{l}{r}-\frac{\Delta_{p}}{2r^2}\right)
\left\{{p_{0(m\phi)}}+\frac{2}{9}
(2{\Pi}_{I0(m\phi)}+{\Pi}_{II0(m\phi)})\right\}\right]\\\label{A14}
&&+\frac{\rho_{0(m\phi)}}{4}\left( 2b'-a'
+\frac{7}{2r}+\frac{6c}{r}-\frac{\Delta_{p}}{r^2}\right),\\\nonumber
&&\bar{E}_{1(N)}=(v_{2(0)N}+w_{2(0)N})\left[\frac{2}{r}
\left[\frac{l}{r}+l'
+\frac{\Delta_{p}}{r^{2}}\right]\right]-(v_{3(0)N}+w_{3(0)N})\frac{r^{2}b}{B_{0}}\\\nonumber
&&-(v_{4(0)N}+w_{4(0)N})\left[c'-\frac{2b}{B_{0}}\right]
+(x_{2(0)N}+y_{2(0)N})\left[\frac{l}{r^{2}}+\frac{l'}{r}
-\frac{\Delta_{p}}{r^{3}}\right]\\\label{A15}
&&+\frac{1}{r}\left[a+2b\right],
\end{eqnarray}
where
\begin{eqnarray}\nonumber
&&\beta_{(N)}=\left((3b+\frac{2c}{r}+\frac{l}{r})
+(v_{1(0)N}+w_{1(0)N})
+r^{2}b(x_{3(0)N}+y_{3(0)N})\right)iw\\\nonumber
&&+(v_{2(0)N}+w_{2(0)N})\left[\frac{l}{r}+b+\frac{2}{r}\right]
+(x_{2(0)N}+y_{2(0)N})\left[\frac{l}{r}+\frac{2b}{B_{0}}\right].
\end{eqnarray}

In the pN approximations, the values of
${\delta_{(m(BD))pN}},~\lambda_{(m(BD))pN}$ and $\bar{E}_{1(pN)}$
are
\begin{eqnarray}\nonumber
&&\delta_{(m(BD))pN}=-\left(1-\frac{2m_0}{r_1}\right)
\left(\frac{p_{0(m\phi)}\beta_{(pN)}}{\rho_{0(m\phi)}+p_{0(m\phi)}}
+\frac{4\Pi_{I0(m\phi)}\beta_{(pN)}}{9(\rho_{0(m\phi)}+\Pi_{I0(m\phi)})}\right.\\\nonumber
&&\left.+\frac{2\Pi_{II0(m\phi)}\beta_{(pN)}}{9(\rho_{0(m\phi)}+\Pi_{II0(m\phi)})}\right)'
-\left\{\frac{p_{0(m\phi)}}{\rho_{0(m\phi)}+p_{0(m\phi)}}
+\frac{4\Pi_{I0(m\phi)}}{9(\rho_{0(m\phi)}+\Pi_{I0(m\phi)})}\right.\\\nonumber
&&\left.+\frac{2\Pi_{II0(m\phi)}}{9(\rho_{0(m\phi)}
+\Pi_{II0(m\phi)})}\right\}\left(1-\frac{2m_0}{r_1}\right){\beta_{(pN)}
}\left\{\frac{7}{4r}+\frac{1}{2}\left(1-\frac{4m_0^2}{r_1^2}\right)\right.\\\nonumber
&&\times\left. \left(1-\frac{m_0^2}{r_1^2}
+\frac{1}{r}\right)\right\}+\frac{1}{2\sqrt{2}r}
\left(1-\frac{m_0}{r_1}\right)\left(1+\frac{3m_0}{r_1}\right)
\left(\frac{\Pi_{k\chi0(m\phi)}\beta_{(pN)}}{\rho_{0(m\phi)}
+\Pi_{k\chi0(m\phi)}}\right)^\theta,\\\label{A16}\\\nonumber
&&\lambda_{(m(BD))pN}=\left[\frac{b\beta_{(pN)}}{2}\left(1-\frac{4m_0}{r_1}\right)\left(1
+\frac{m_0}{r_1}\right)\left(1-\frac{3m_0}{r_1}\right)\left(1
-\frac{m_0}{r_1}\right)'\right.\\\nonumber
&&\times\left\{{p_{0(m\phi)}}'+\frac{2}{9}(2{\Pi}_{I0(m\phi)}'+{\Pi}_{II0(m\phi)}')\right\}-
\left[\left\{\left(\frac{c}{r}\right)'+\frac{3}{4r}
\left(\frac{l}{r}+{l'}-\frac{\Delta_{p}}{2r^2}\right)
\right\}\right.\\\nonumber
&&+\frac{1}{2r^2}\left(1-\frac{4m_0^2}{r_1^2}\right)\left(
{2a}\left(1+\frac{m_0}{r_1}\right)+{2b}
\left(1-\frac{m_0}{r_1}\right)
-\frac{\Delta_{p}}{2r^2}\right)\left(\left(1-\frac{m_0}{r_1}\right)'\right.\\\nonumber
&&\times\left.\left(1+\frac{m_0}{r_1}\right)+\frac{1}{r}
\right)+\frac{1}{2}\left.\left({a}\left(1+\frac{m_0}{r_1}\right)
+{b}\left(1-\frac{m_0}{r_1}\right)\right)'\right]\left\{{p_{0(m\phi)}}\right.\\\nonumber
&&\left.+\frac{2}{9} (2{\Pi}_{I0(m\phi)}+{\Pi}_{II0(m\phi)})\right\}
\left(1-\frac{2m_0}{r_1}\right)-{2b}\left(1-\frac{2m_0}{r_1}\right)
\left\{{p_{0(m\phi)}}\right.\\\nonumber
&&\left.+\frac{2}{9}(2{\Pi}_{I0(m\phi)}+{\Pi}_{II0(m\phi)})\right\}
\left\{\frac{7}{4r}\right.+\frac{1}{2}\left(1-\frac{4m_0^2}{r_1^2}\right)
\left(\left(1-\frac{m_0}{r_1}\right)'\right.\\\nonumber
&&\left.\times\left(1 +\frac{m_0}{r_1}\right)\right.
\left.\left.+\frac{1}{r}\right)\right\}+\frac{1}{2\sqrt{2}r}
{\Pi}_{k\chi0(m\phi)}^{\theta}\left(1-\frac{m_0}{r_1}\right)
\left(1+\frac{3m_0}{r_1}\right)\\\nonumber &&\times
\left({a}\left(1+\frac{m_0}{r_1}\right)+{3b}\left(1
-\frac{m_0}{r_1}\right)-\frac{3\delta_{p}}{2r^2}\right)
+\frac{\Pi_{k\chi0(m\phi)}}{2\sqrt{2}r}\left[6\left(1-\frac{m_0}{r_1}
\right)\right.\\\nonumber
&&\times\left.\left(1+\frac{m_0}{r_1}\right)^\theta
\left({b^\theta}\left(1-\frac{m_0}{r_1}\right)^\theta
+{b}\left(1-\frac{m_0}{r_1}\right)\right)\right.\left({a}
\left(1+\frac{m_0}{r_1}\right) +\frac{c}{r}\right)^\theta\\\nonumber
&&+\left.2\left(1-\frac{4m_0^2}{r_1^2}\right)
+\left({2a}\left(1+\frac{m_0}{r_1}\right)+{2b}\left(1
-\frac{m_0}{r_1}\right)-\frac{\delta_{p}}{r^2}\right)\right.\\\nonumber&&\left.\times
\left({a}\left(1+\frac{m_0}{r_1}\right)+{b}\left(1-\frac{m_0}{r_1}\right)\right)^\theta\right]
-\frac{\rho_{0(m\phi)}}{4}\left(1-\frac{4m_0}{r_1}\right)\\\nonumber
&&\times\left\{\left(
{a}\left(1+\frac{m_0}{r_1}\right)\right)'-\frac{1}{r}\left(1
-\frac{m_0}{r_1}\right)^\theta
\left(1+\frac{m_0}{r_1}\right)\left(1-\frac{2m_0}{r_1}\right)\right.\\\nonumber
&&\times\left.\left(\frac{l}{r}+{a^{\theta}}\left(1+\frac{m_0}{r_1}\right)^\theta
-{a}\left(1+\frac{m_0}{r_1}\right)-{2b}\left(1-\frac{m_0}{r_1}\right)\right)\right\}
+\frac{\rho_{0(m\phi)}}{2\sqrt{2}r}\\\nonumber
&&\times\left(1-\frac{2m_0}{r_1}\right)\left(\frac{2l}{r}+{2a}
\left(1+\frac{m_0}{r_1}\right)-\frac{\Delta_{p}}{r^2}\right)
+\frac{\mu_0}{2\sqrt{2}}\left(1-\frac{2m_0}{r_1}\right)\left(\frac{l}{r}+{b}\right.\\\nonumber
&&\left.\times\left(1-\frac{m_0}{r_1}\right)\right)'
\left.-\frac{\beta_{p}}{2\sqrt{2}}\left(1-\frac{2m_0}{r_1}\right)
\right.\left.\left\{\frac{3}{2r}+\left(1+\frac{m_0}{r_1}\right)'
\left(1-\frac{m_0}{r_1}\right)\right\}\right],\\\label{A17}\\\nonumber
&&\bar{E}_{1(pN)}=\left[(\dot{x}_{1(ap)}^{(pN)}+\dot{y}_{1(ap)}^{(pN)})
+(v_{2(ap)}^{'(pN)}+w_{2(ap)}^{'(pN)})+(x_{2(ap)}^{\theta(pN)}+y_{2(ap)}^{\theta(pN)})\right.\\\nonumber
&&\left.+(v_{1(ap)}^{(pN)}+w_{1(ap)}^{(pN)})(\frac{m_{0}}{r})'+(v_{2(ap)}^{(pN)}+w_{2(ap)}^{(pN)})
\left[4(\frac{m_{0}}{r})'+\frac{l}{r}\right]\right]iw+(v_{2(bp)}^{(pN)}\\\nonumber
&&+w_{2(bp)}^{(pN)})'+(x_{2(ap)}^{(pN)}
+y_{2(ap)}^{(pN)})^{\theta}+(v_{1(0p)}^{(pN)}+w_{2(0p)}^{(pN)})(\frac{m_{0}}{r})',\\\label{A18}
\end{eqnarray}
where
\begin{equation}\nonumber
\beta_{(pN)}=\left((F_{(m\phi)pN}+\bar{E}_{0(a)pN})iw+\bar{E}_{0(b)pN}\right)
\end{equation}

\end{document}